\documentclass[twocolumn,showpacs]{revtex4}
\usepackage{graphics}
\usepackage{epsfig}
\usepackage{graphicx}
\usepackage{epstopdf}
\usepackage{bm}

\begin{document}
\title{Rydberg atoms in one-dimensional optical lattices}
\author{M. Viteau$^1$, M. G. Bason$^1$, J. Radogostowicz$^{2,3}$, N. Malossi$^3$, D. Ciampini$^{1,2,3}$, O. Morsch$^{1*}$, E. Arimondo$^{1,2,3}$}
\affiliation{$^1$INO-CNR, Dipartimento di Fisica `E. Fermi',$^2$Dipartimento di Fisica `E. Fermi', Universit\`a di
Pisa, $^3$CNISM UdR, Dipartimento di Fisica `E. Fermi', Universit\`a di
Pisa, Largo Pontecorvo 3, 56127 Pisa, Italy}

\begin{abstract}
We experimentally realize Rydberg excitations in Bose-Einstein condensates of rubidium atoms loaded into quasi one-dimensional traps and in optical lattices. Our results for condensates expanded to different sizes in the one-dimensional trap agree well with the intuitive picture of a chain of Rydberg excitations. We also find that the Rydberg excitations in the optical lattice do not destroy the phase coherence of the condensate, and our results in that system agree with the picture of localized collective Rydberg excitations including nearest-neighbour blockade.
\end{abstract}

\pacs{03.65.Xp, 03.75.Lm}
\maketitle

Rapid progress in cold atom physics in recent years has led to ambitious proposals in which the strong and controllable long-range interactions between Rydberg atoms \cite{comparat_2010} are used in order to implement quantum computation schemes \cite{jaksch_2000,lukin_2001,muller_2009,isenhower_2010,wilk_2010} and quantum simulators \cite{ryabtsev_2005,buluta_2009,weimer_2010}. One central ingredient of these schemes is the dipole blockade, which has been observed for pairs of atoms \cite{gaetan_2009,urban_2009} and disordered ultra-cold atomic clouds \cite{tong_2004,carroll_2006,reetzlamour_2008,heidemann_2007,heidemann_2008,low_2009}. In the dipole blockade mechanism the excitation of an atom to a Rydberg state is inhibited if another, already excited, atom is less than the so-called blockade radius $r_b$ away \cite{comparat_2010}.\\
\indent  In the case of a large number of atoms, a simple physical picture is that of a collection of super-atoms \cite{vuletic_2006} in which a single excitation is shared among all the atoms inside the blockade radius \cite{robicheaux_2005,heidemann_2007,stanojevic_2009}. As pointed out in~\cite{lukin_2001}, this suggests that experiments requiring single atoms excited to Rydberg states can also be performed with such collective Rydberg excitations, each containing several hundreds or thousand of atoms  making it possible to work with, e.g., Bose condensates loaded into optical lattices without the need to induce a Mott insulator transition in order to have single-atom occupation of the lattice sites. \\
\indent Here we demonstrate the realization of one-dimensional chains of collective Rydberg excitations in rubidium Bose condensates. In addition we study the excitation dynamics of up to 50 Rydberg excitations in a condensate occupying around 100 sites of a one-dimensional optical lattice, the excitation number determined by the ratio between blockade radius and lattice spacing. Classical and quantum correlations and highly entangled  collective states are  expected to be created, as pointed out in~\cite{lesanovsky_2011} for one dimensional Rydberg gases and in~\cite{zuo_2010} for one-dimensional optical lattices. Our results pave the way towards their controlled creation. Our experiments make it necessary to combine techniques for obtaining Bose-Einstein condensates in dipole traps and optical lattices with those for creating and detecting Rydberg atoms. While the former require a vacuum apparatus with large optical access, the latter demand precisely controlled electric fields. In our setup (see Fig. 1), we reconciled these requirements by adding external field electrodes to the quartz cell of our cold atom apparatus and carefully studying the operating conditions under which the best field control and highest detection efficiency of the Rydberg atoms was achieved.

\begin{figure}[htp]
\includegraphics[width=8cm]{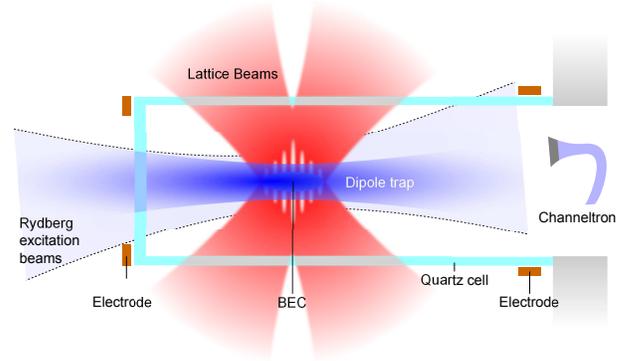}
\caption{Experimental setup. The Bose-Einstein condensates are created in the optical dipole trap (only one arm of the trap is shown here) and can then be expanded in the dipole trap by switching off one of its arms and / or loaded into the optical lattice. The Rydberg excitation beams are almost parallel to the horizontal arm of the dipole trap, and field ionization and detection of the Rydberg excitations is achieved using the electrodes (outside the quartz cell) and the channeltron, respectively.}
\end{figure}

In our experiments we first create Bose-Einstein condensates of up to $10^5$ 87-Rb atoms using a two-step evaporation protocol with a TOP-trap and a crossed optical dipole trap. Quasi one-dimensional atomic samples of length $l$ are then created by switching off one of the trap beams and allowing the condensate to expand inside the (now one-dimensional) dipole trap for variable times (up to $500\,\mathrm{ms}$). Rydberg states with principal quantum number $n$ between $55$ and $80$ are subsequently excited using a two-colour coherent excitation scheme with one laser at $420\,\mathrm{nm}$ (obtained by frequency doubling of a $840\,\mathrm{nm}$ MOPA beam; beam waist $120\,\mathrm{\mu m}$, maximum power $20\,\mathrm{mW}$) blue-detuned by $0.5-1\,\mathrm{GHz}$ from the $6P_\mathrm{3/2}$ energy level of 87-Rb, and a second laser at $1010-1030\,\mathrm{nm}$ (beam waist $150\,\mathrm{\mu m}$, maximum power $150\,\mathrm{mW}$). Both lasers are locked via a Fabry-Perot cavity to a stabilized reference laser at $780\,\mathrm{nm}$. Finally, detection of the Rydberg atoms is achieved with efficiency $\eta =35\pm 10\%$ \cite{viteau_2010} by field ionization followed by a two-stage acceleration towards a channeltron charge multiplier. The electrodes for the field ionization and the acceleration are placed outside the glass cell in which the condensate is created. In previous experiments using Rydberg states with $n=30-85$ in various electric fields we demonstrated a control of the electric field at the level of a few mV cm$^{-1}$~\cite{viteau_2011}.\\
\indent The highly elongated atomic clouds created by the method just described are up to $l=1\,\mathrm{mm}$ long, while their radial dimensions are on the order of $1-2\,\mathrm{\mu m}$ (radial dipole trap frequencies are around $100\,\mathrm{Hz}$). Since the expected blockade radii for the Rydberg states between $n=50$ and $n=80$ used in our experiments range from $5\,\mathrm{\mu m}$ to $15\,\mathrm{\mu m}$, this suggests that at most one Rydberg excitation fits radially into the condensate and that the total number of Rydberg excitations inside such a sample should depend on $l$. In order to test this picture we align the two Rydberg excitation lasers so as to be almost parallel to the dipole trap beam in which the condensate expands. After expanding the condensates, Rydberg atoms are created using pulses of up to $1.5\,\mathrm{\mu s}$ duration (during which the condensate expansion is frozen and Penning and blackbody ionization \cite{saffman_2008} were found to be negligible) and finally detected by field ionization. The duration of the excitation pulses is chosen such that the system is in the saturated regime in which the number of Rydberg atoms levels off after an initial increase on a timescale of hundreds of nanoseconds \cite{heidemann_2007}. Figure 2(a) shows typical results of such an experiment for excitation to the $66D_{5/2}$ state using a $1\,\mathrm{\mu s}$ excitation pulse, in which a linear increase of the number of Rydberg atoms with the length of the condensate is visible. This result agrees with the simple intuitive picture of super-atoms being stacked in a one-dimensional array of varying length. The highly correlated character of the Rydberg excitations thus created is further demonstrated by analyzing the counting statistics of our experiments, which are strongly sub-Poissonian for short expansion times of a few milliseconds with observed Mandel $Q$-factors \cite{ates_2006}of $Q_{obs}\approx -0.3$, where $Q_{obs}=\Delta N/\langle N\rangle-1$ with $\Delta N$ and $\langle N\rangle$ the standard deviation and mean of the counting statistics, respectively. Taking into account the detection efficiency $\eta$ for Rydberg excitations in our system, this indicates actual $Q$-factors close to $-1$ (since the observed $Q$-factor $Q_\mathrm{obs}=\eta Q_\mathrm{act}$).
\begin{figure}[htp]
\includegraphics[width=7cm]{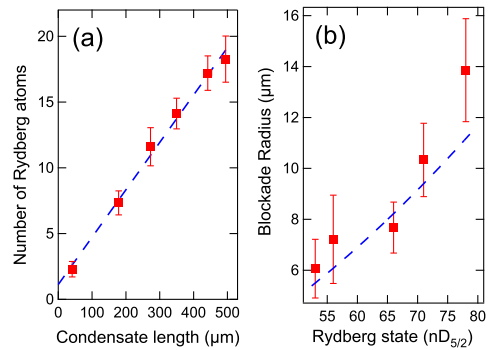}
\caption{Rydberg excitation in an expanded condensate. (a) Number of $66D_{5/2}$-Rydberg excitations (derived from the number of detected ions and the detector efficiency) for an excitation pulse of $1\,\mathrm{\mu s}$ duration as a function of the condensate length. The error bars indicate the standard deviation of the mean. (b) Measured blockade radius $r_b$ as a function of the principal quantum number $n$. The dashed line is the theoretically predicted value assuming a total laser linewidth of $300\,\mathrm{kHz}$.}
\end{figure}
\indent From the data in Fig. 2(a) we can extract the mean distance between adjacent super-atoms, assuming a close-packed filling of the one-dimensional atomic cloud. To this end we take into account that in the saturated regime the individual super-atoms perform Rabi oscillations between their ground and excited states and that hence on average only half of the super-atoms are in the excited state. In Fig. 2(b) the blockade radius measured in this way is shown for different Rydberg states, together with the theoretically expected value \cite{comparat_2010} for a pure van-der-Waals interaction following an $n^{11/6}$ scaling (due to the $n^{11}$ scaling of the $C_6$ coefficient and the $r^{-6}$ dependence of the interaction on the atomic distance $r$) and assuming a total linewidth (intrinsic plus Fourier-limited) of our excitation lasers of around $300\,\mathrm{kHz}$. The measured values for the blockade radii of $5-15\,\mathrm{\mu m}$ confirm our interpretation of an essentially one-dimensional chain of Rydberg excitations. For high-lying Rydberg states with $n>75$ we find deviations from the theoretically expected scaling with $n$, which may be due to small electric background fields. By deliberately applying a small electric field during the excitation pulse we have checked that for large $n$, an electric field on the order of a few $\mathrm{mV/cm}$ can lead to a dipole-dipole interaction energy contribution (scaling as $n^{4/3}$) on the order of the usual van-der-Waals interaction, resulting in an effective blockade radius that is substantially larger than what is expected for the pure van-der-Waals case. These observations confirm that in spite of the compromise made in our setup in order to accommodate the necessary laser beams as well the electric field plates and the channeltron, we are able to avoid background electric fields larger than around $1\,\mathrm{mV/cm}$ by carefully choosing the parameters for our pulse sequences.\\
\begin{figure}[htp]
\includegraphics[width=5cm]{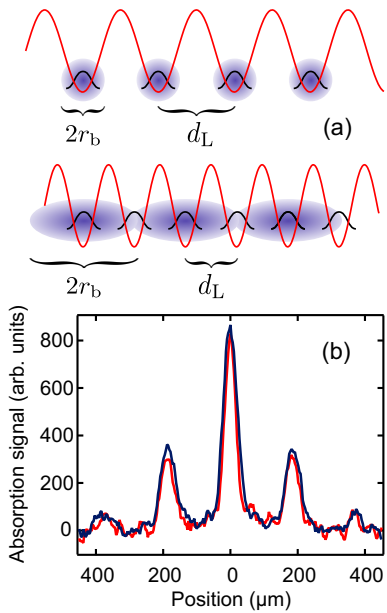}
\caption{Rydberg excitation in an optical lattice.
(a) Schematic representation of the experiment. Depending on the lattice spacing $d_L$ and the blockade radius $r_b$, a collective Rydberg excitation (indicated by the blue spheres) can either be confined to a single lattice site (above) or extend over several lattices sites (below). (b) Time-of-flight profile of a condensate released from an optical lattice with $d_L=0.42\,\mathrm{\mu m}$, $V_0/E_\mathrm{rec}\sim 10$ and $r_b\approx 6\,\mathrm{\mu m}$ before (blue line) and after ten cycles of Rydberg excitation to the $53D_{5/2}$ state (pulse duration $1\,\mathrm{\mu s}$) and subsequent detection of the Rydberg atoms by field ionization (red line). The virtually identical interference profiles show that the overall phase coherence of the condensate is not affected by the collective Rydberg excitations.}
\end{figure}
\indent Having demonstrated the one-dimensional character of the Rydberg excitation in the elongated condensate, we now add an optical lattice along the direction of the one-dimensional sample. by intersecting two linearly polarized laser beams of wavelength $\lambda_L=840\,\mathrm{nm}$ at an angle $\theta$, leading to a lattice with spacing $d_L=\lambda_L/(2\sin(\theta/2))$. The optical access in our setup allows us to realize spacings up to $\approx 13\,\mathrm{\mu m}$. For the smallest lattice spacing $d_L=\lambda_L/2=0.42\,\mathrm{\mu m}$, the maximum depth $V_0$ of the periodic potential $V(x)=V_0\sin(2\pi x/d_L)$ measured in units of the recoil energy $E_\mathrm{rec}=\hbar^2\pi^2/(2m d_L^2)$ (where $m$ is the mass of the atoms) is $V_0/E_\mathrm{rec}\sim 20$, whereas for the largest lattice spacing the lattice depth can be up to $V_0/E_\mathrm{rec}\sim 5000$.  By varying the lattice spacing between $d_L=0.42\,\mathrm{\mu m}$ and $\approx 13\,\mathrm{\mu m}$ we interpolate between the extremes $d_L\ll r_b$ in which the Rydberg excitation on one site interacts strongly with other excitations many sites away, and $d_L\gtrsim r_b$ where only nearest neighbours are expected to interact significantly (see Fig 3a). Since in the lattice direction the size $\Delta x$ of the on-site wavefunction is around $\Delta x = d_L/(\pi(V_0/E_\mathrm{rec})^{1/4})$, the depth of the optical lattice is chosen to ensure that the ratio between the size $\Delta x$ of an on-site wavefunction and $d_L$ is less than $\approx 0.1$. Therefore the spatial extension of the condensate wavefunction, highly localized at each lattice site, is much less than $r_b$.

In a preliminary experiment we test the effect of several Rydberg excitation cycles on the condensate phase coherence between the lattice sites for a lattice spacing $d_L=0.42\,\mathrm{\mu m}$ and hence in the limit $d_L\ll r_b$. After creating Rydberg excitations with a $1\,\mathrm{\mu s}$ pulse we perform a time-of-flight experiment by suddenly switching off the dipole trap and the optical lattice and imaging the atomic distribution after $23\,\mathrm{ms}$ of free fall. The resulting interference pattern can be used to confirm that, even after the projective Rydberg measurement by field ionization, the degree of phase coherence of the ground-state condensate wavefunction is not modified by the short-range correlations among the atoms created by the collective Rydberg excitations. As shown in Fig. 3b, more than $10$ excitation-detection cycles can be performed on the same condensate without a noticeable loss of phase coherence or atom number, allowing us to enhance the statistical sample size in our experiments.

\begin{figure}[htp]
\includegraphics[width=5cm]{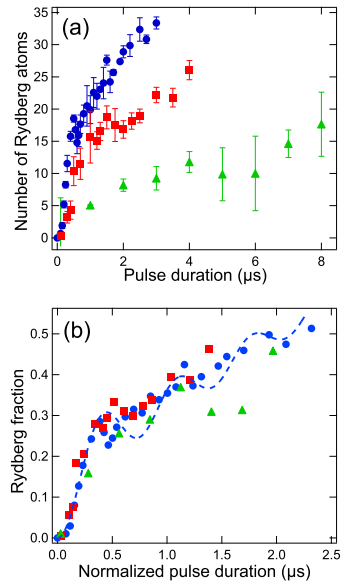}
\caption{Dynamics of Rydberg excitations in an optical lattice. (a) Number of $53D_{5/2}$ Rydberg excitations as a function of  the $\tau_p$ pulse duration for different average atom numbers per lattice site: $\langle N_i\rangle=500$ (filled circles), $\langle N_i\rangle=200$ (open squares) and $\langle N_i\rangle=50$ (filled triangles). (b) Here the experimental results of (a) are plotted against the renormalized pulse duration $\tau\sqrt{\langle N_{max}\rangle /\langle N_i\rangle}$ with $\langle N_{max}\rangle$ the largest atom number. The vertical axis is scaled in terms of the fraction of lattice sites containing a collective excitation. For clarity, in (b) error bars have been omitted. The dashed line is the numerical simulation for $\alpha=2$ (see text).}
\end{figure}

In order to study the dynamics of Rydberg excitations in the lattice in the regime $d_L\approx r_b$, we create a one-dimensional sample of well-defined length with a reasonably uniform atom distribution, sharp boundaries and a controllable average atom number per lattice site by expanding the condensate for a given time, then ramping up the power of the lattice beams and ramping down the power of the dipole trap beam to a value for which outside the lattice region atoms are lost from the trap. In the overlap region of the lattice and the dipole trap (whose length is given roughly by the lattice beam diameter) the combined radial trap depth is sufficient to hold the atoms against gravity. The $n_L\approx 100$ filled lattice sites ($d_L=2.27\,\mathrm{\mu m}$ is chosen for this experiment) each contain on average $N_i=50-500$ atoms. Since the dimensions of an on-site wavefunction are around $2\,\mathrm{\mu m}$ radially and $0.2\,\mathrm{\mu m}$ in the lattice direction, we expect each lattice site to contain at most one single collective Rydberg excitation. In this regime, then, we effectively have `zero-dimensional' atomic clouds on each lattice site. In this system we measure the number of atoms in the $53D_{5/2}$ Rydberg state as a function of time by varying the duration of the excitation pulse (see Fig. 4a). Our observations can be interpreted using a simple numerical model that contains a truncated gaussian distribution $N_i$ of atom numbers on the lattice sites labeled by $i$ leading to a distribution of on-site collective Rabi frequencies $\sqrt{\alpha N_i}\Omega_\mathrm{single}$, where the single-particle Rabi frequency $\Omega_\mathrm{single}\sim 2\pi\times 30\,\mathrm{kHz}$ for our experimental parameters. Here $\alpha=[r_b/d_L]$ is the nearest (larger) integer number to $r_b/d_L$ and describes the number of sites taking part in a collective excitation. Another way of putting this is to observe that since $\alpha$ Rydberg excitations of neighbouring lattice sites are suppressed by the dipole blockade, on average, only $1/\alpha$ of the lattice sites can contain an excitation. (The interactions between more distant neighbours through the van-der-Waals interaction are smaller by a factor of $2^6$ compared to nearest-neighbour interactions.) Taking into account the average over the Rabi-flopping cycle of each excitation, when $\alpha=2$ we expect to have around $\frac{1}{4}n_L$ collective excitations present in the lattice in the saturated regime.

In order to visualize the scaling of the collective Rabi frequency with $\sqrt{N_i}$, in Fig. 4b we plot the results of our experiments as a function of the pulse duration normalized to the collective Rabi frequency for the largest $N_i$. Using that scaling, all the curves of Fig. 4a collapse onto a single curve. One also sees a distinct change in the slope of the excitation curves at a normalized pulse duration of $0.5\,\mathrm{\mu s}$, where the fraction of lattice sites containing a Rydberg excitation is around $0.25$. This point indicates the crossover to the saturated regime in which the number of excitations should, in principle, level out but in practice continues to grow. We believe this slow increase for long excitation times to be due to dephasing effects induced by a coupling between the super-atoms. That coupling can be introduced into our model via a small dephasing rate $\delta$ between the lattice sites, chosen such as to reproduce the observed long-term increase in the number of collective excitations. Finally, in the transient regime we see the remnants of a coherent Rabi oscillation, which is washed out due to the different local collective Rabi oscillations adding up. The expected number of Rydberg excitations as a function of time $P(t)$ is then given by
\begin{equation}
P(t)=\frac{1}{\alpha}\sum_i\sin^2\left(\sqrt{\alpha N_i}\Omega_\mathrm{single}t+\delta N_\mathrm{tot}t\right),
\end{equation}
where $N_\mathrm{tot}=\sum_i N_i$ is the total number of atoms and the best fit is found for $\delta\approx 10^{-5}$. The factor $\alpha$ outside the summation accounts for multiple counting when $\alpha>1$. As can be seen in Fig. 4b, this model reproduces the basic features observed in our experiment.\\
\indent We have demonstrated the controlled preparation of Rydberg excitations in large ensembles of ultracold atoms in one-dimensional dipole traps and in optical lattices. Our results can straightforwardly be extended to two- and three-dimensional lattice geometries, and the lattice spacings realizable in this work should allow single-site addressing as well as selective control of the Rydberg excitations. In particular, it will be important  to verify whether long-range crystalline order can be created by implementing adiabatic transfer protocols as recently proposed in \cite{pohl_2010,schachenmayer_2010}. Furthermore, appropriate ion detection techniques such as those involving microchannel plates should allow direct observation of the distribution of Rydberg excitations in the lattice and in the one-dimensional trapping geometry.\\
\indent Financial support by the E.U.-STREP ``NAMEQUAM'', the MIUR PRIN-2007 Project and by a CNISM ``Progetto Innesco 2007'' is gratefully acknowledged. We thank I. Lesanovsky, T. Pohl, P. Pillet, D. Comparat and G. Morigi for stimulating discussions, S. Rolston and R. Fazio for a thorough reading of the manuscript, and P. Huillery for assistance.

\bibliographystyle{apsrmp}

\newpage

\end{document}